\def\year{2022}\relax
\documentclass[letterpaper]{article} 
\usepackage{aaai21}  
\usepackage{times}  
\usepackage{helvet} 
\usepackage{courier}  
\usepackage[hyphens]{url}  
\usepackage{graphicx} 
\usepackage{natbib}  
\usepackage{caption} 
\usepackage{graphicx,color}
\usepackage{amsmath, array}
\usepackage{placeins}
\usepackage{changepage}
\usepackage{longtable}
\usepackage{multirow}
\usepackage{siunitx}
\usepackage{supertabular}
\usepackage{enumitem}
 \usepackage{textcomp} 
\usepackage{url} 
\usepackage{array}

\usepackage{subcaption}
\usepackage{comment}
\usepackage{verbatim}

\title{Characterizing Retweet Bots: The Case of Black Market Accounts}

\title{Characterizing Retweet Bots: The Case of Black Market Accounts}
\author {
    Tuğrulcan Elmas,
    Rebekah Overdorf, 
    Karl Aberer \\
}
\affiliations {
    EPFL \\
    tugrulcan.elmas@epfl.ch, 
    rebekah.overdorf@epfl.ch, 
    karl.aberer@epfl.ch
}

\begin{document}

\newcommand{\Secref}[1]{Section~\ref{#1}}
\newcommand{\Figref}[1]{Fig.~\ref{#1}}
\newcommand{\Tabref}[1]{Table~\ref{#1}}
\newcommand{\archive}{archive}
\newcommand{\Archive}{Archive}
\newcommand{\integrity}{elections-integrity }
\newcommand{\Integrity}{Elections-integrity }
\newcommand{\snc}{screenname-changes }
\newcommand{\name}{ephemeral \xspace} 
\newcommand{\Name}{Ephemeral \xspace}
\newcommand{\descr}[1]{{\bigskip\noindent\textbf{#1}}}
\newcommand{\descrplain}[1]{{\smallskip\noindent\textbf{#1}}}

\newcommand{\attr}[1]{#1}

\newcommand{\bekah}[1]{{\color{blue}BO: #1}}


\maketitle
\begin{abstract}

Malicious Twitter bots are detrimental to public discourse on social media. Past studies have looked at spammers, fake followers, and astroturfing bots, but retweet bots, which artificially inflate content, are not well understood. We  present the first study focusing exclusively on retweet bots. We characterize retweet bots that have been uncovered by purchasing retweets from the black market. We detect whether they are fake or genuine accounts involved in inauthentic activities and what they do in order to appear legitimate. We also analyze their differences from human-controlled accounts. From our findings on the nature and life-cycle of retweet bots, we also point out several inconsistencies between the retweet bots used in this work and bots studied in prior works. Our findings challenge some of the fundamental assumptions related to bots and in particular how to detect them.

\end{abstract}
\section{Introduction}

An extensive amount of research has focused on detecting automated accounts, \emph{bots}, on Twitter. While some studies directly address bots' functions, such as spamming~\cite{yardi2010detecting}, pushing slogans on top of Twitter trends~\cite{elmas2020power}, and inflating follower counts~\cite{fameforsale}, others use the general term \emph{social bot} to mean \emph{accounts that mimic humans} without addressing how the accounts are automated. By only considering generic social bots and their detection, we cannot learn about the actual nature of the bots themselves.

Notions about the ``lives'' of retweet bots, i.e., bots that retweet posts that others pay them to promote, have yet to be explored. While there are studies that detect coordinated groups of accounts, of which retweet bots may be a part, there is no work that tries to understand what a single retweet bot looks like, how it behaves, and how it differs from a human account.  If a user is a dedicated fan of a football club and, thus, uses their account only to retweet that club's Twitter posts, what makes them different from a retweet bot? 

Understanding the nature of retweet bots is challenging in the absence of reliable ground truth on retweet bots. The standard method for labeling bots, annotating retweet bots by hand, is inexact. As illustrated by the football club fan, determining whether an account is controlled by an algorithm or a human is not straightforward. Using human annotation is error-prone and can lead to results that are biased by the annotator's assumptions about what a bot is. 

We address this by studying retweet bots whose services were directly purchased by prior work. Using this dataset of reliable retweet bots, we also examine both the assumptions and findings made by prior works and find some that do not hold up against the bots in this dataset. This allows us to observe inconsistencies in how bots behave. We focus on four research questions:

\begin{enumerate}[label=\textbf{{RQ{\theenumi}}},leftmargin=26pt]
    \item Where do these retweet bots come from? Were they created for this purpose or compromised?
    \item What is the lifetime of these retweet bots?
    \item How do the retweeters in our dataset act differently from human users?
    \item Are there any differences between the bots examined in this work and those found in prior studies? 
\end{enumerate}

In answering these questions, we 1) present the first study focusing on retweet bots exclusively; 2) characterize retweet bots, providing evidence that some are mass-created and controlled by one center entity while some are compromised and used aggressively; 3) challenge fundamental assumptions about the nature of bot accounts, such as account age and over-activity; and 4) we discuss challenges in bot detection with respect to retweet bots.

\section{Background and Related Work}
\label{sec:related}

The literature on the detection and analysis of bots principally defines and annotates bots either by their \emph{nature} or their \emph{primary function}. The popular term ``social bot'' in reference to ``bots mimicking human behavior'' is an example of the former~\cite{boshmaf2011socialbot,ferrara2016rise}, which are usually reliant on human annotation, which we know to be unreliable~\cite{cresciinferhumans}. The latter, e.g., spammers~\cite{spamcampaign,yardi2010detecting,herzallah2018feature}, fake followers~\cite{fameforsale}, and astroturfing bots~\cite{elmas2020power}, are usually less reliant on human annotation since the function of an account is more straightforward to define and detect based on specific behavior. The bots we focus on in this work are distinguished by their primary \emph{function}: retweeting other accounts. We will refer to these accounts as \emph{retweet bots}.

In order to forgo error-prone machine or human-based detection, some studies~\cite{fameforsale} opt for the more reliable method of directly purchasing a bot service. Past work has shown that this method can be used to study bots that inflate follower counts and to analyze the authors of such posts~\cite{dutta2021abome,collusiveretweeters}. We follow this more reliable method of bot collection and analyze the \emph{accounts} controlled by a vendor (or vendors) who sells retweets.

Prior work on retweet bots is not devoid of research on accounts, however, these studies focus on coordinated groups of accounts~\cite{liu2017holoscope,vargas2020detection}, including those who retweet others~\cite{gupta2019malreg,mazza2019rtbust}, and not on individual accounts. By studying individual accounts we learn how the accounts became retweeters, how long they were active, and how they were different from genuine accounts. Such studies are also constrained by a single topic (e.g., finance~\cite{cresci2019cashtag}) or by an assumption (e.g., that bots always act at roughly the same time~\cite{debot} or that bots' timelines are similar~\cite{cresci2017social}). 

To the best of our knowledge, there are only two prior works that provide a per-account analysis of retweet bots. Unlike the data presented in this paper, both rely on human annotation. Dutta et al. (\citeyear{dutta2020hawkeseye}) took a much more restrictive definition of \emph{retweeter}, effectively only studying trend spammers and not a broader population of accounts. They also leverage human annotation which partially relies on whether or not a large number of tweets/retweets were posted within a short time period, something we find is not the case with the retweet bots in our dataset. Giatsogloue et al. (\citeyear{giatsoglou2015retweeting}) studied both retweeted posts and retweeters and proposed a retweeter classification method. The overlapping observations between our work and this are that 1) retweeters have a high follower friend ratio, in contrast to popular belief that they do not, and 2) they retweet with similar time delays. We expand on these observations and explore further.
\section{Dataset Overview}
\label{sec:dataset}

We use two distinct types of data for this analysis: 1) data from retweet bots and 2) data from human (genuine) accounts which we use as a control group. 

\subsubsection{Retweet Bots}
    
The retweet bot dataset is made up of bots that retweet others' tweets. The dataset was introduced in a paper by Golbeck (\citeyear{golbeck2019benford}). In this work, the author created a fake and ``uninteresting'' Twitter account with no followers and posted ``uninteresting'' tweets in order not to attract genuine retweeters to the tweets. They then contacted vendors selling ``retweet services'' and purchased 100 retweets for each post. The goal of this study was to detect whether a post had been promoted by retweet fraud or genuinely received the retweets. We build on this work by exploring the accounts that participated in the retweeting activity. 

To build this dataset, we started with the 18 ``uninteresting'' tweets included in the Golbeck study and collected all of the accounts that retweeted any of these tweets. Due to the 3.5 year gap between the original study and this study, most accounts were either suspended or inactive, leaving only 862 non-suspended accounts. Although these accounts were not suspended and therefore Twitter has not flagged them as bots, because the activities (retweets) of these bots were directly purchased, we can be certain that these 862 accounts were in fact acting as retweet bots during the time of Golbeck's study.

To extend this dataset, Golbeck extended the search for accounts laterally. That is, for each bot $A$ that retweeted one of the ``uninteresting'' tweets, find all of the other posts, e.g., $t$, that $A$ retweeted. Because $A$ only retweets posts they are paid to retweet, we know that the author of $t$ paid for $A$ and other bots to retweet it. Therefore, we can reasonably assume that \emph{some} of the other accounts retweeting $t$ are also retweet bots and they are recorded as such. This yielded a dataset of 6,112 accounts, 5,332 of which were suspended. We discarded the 780 non-suspended accounts from this study since there is some uncertainty in this collection method and, at least so far, Twitter does not suspect that these accounts are fake, so they may be genuine. We keep the suspended accounts since Twitter has more or less corroborated these results. We do not take for granted that these are all retweet bots. We explore later whether the suspended accounts are genuine or were suspended for some other reason, but find that this is very unlikely the case. Our final dataset consists of 6,199 accounts. Most of the analysis in this paper is focused on the 862 accounts whose actions were directly purchased and were not suspended (so we have the full data from each account). 

Using the Twitter API, we collected the timeline (most recent 3,200 tweets) from all 862 non-suspended accounts in October 2020. We refer to this dataset as the \textit{timeline} dataset. This dataset consists of 1,212,030 retweets and 125,974 tweets. Some of the bots in the timeline dataset have few retweets, e.g. 48 bots have less than 100 retweets. One likely explanation is that there were more retweets, but they have since been deleted. We include all of the accounts in the dataset and the analysis regardless of retweet count. 

\begin{figure*}[htb]
    \centering
    \includegraphics[width=0.9\textwidth]{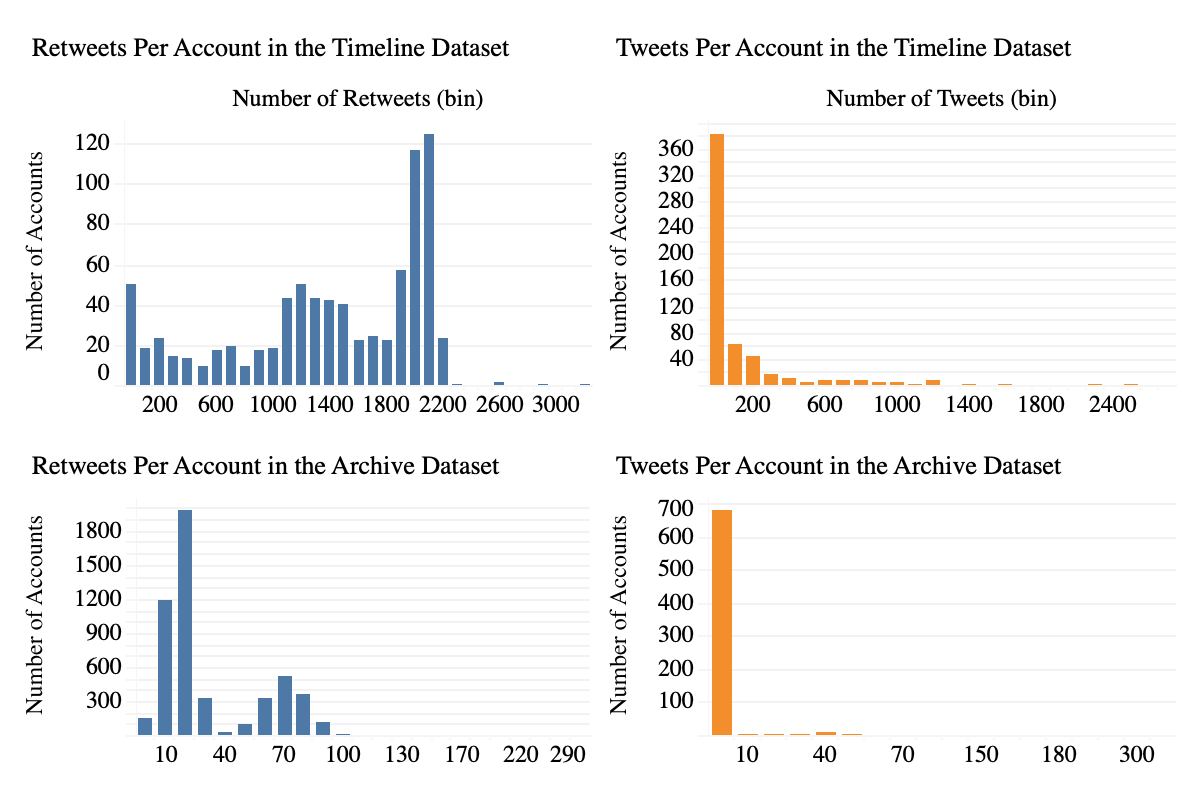}
    \caption{Number of retweets (left) and the tweets (right) per account in the timeline dataset (top) and the archive dataset (bottom). As the timeline dataset is collected using Twitter API, it consists way more tweets and retweets than the archive dataset. However, there are more accounts in the archive dataset. }
    \label{fig:descriptive}
\end{figure*}

Collecting the data from the 5,332 suspended accounts was more challenging. Internet archive's Twitter Stream Grab provides 1\% of all tweets since 2011~\cite{team2020archive} and has been used extensively by past research~\cite{tekumalla2020mining,elmas2021dataset, elmas2020misleading}. By mining this dataset, we collected roughly 1\% of all tweets from these accounts and their profile information. We call this dataset the \textit{archive} dataset. This dataset consists of 301,932 retweets and 29,899 tweets. \Figref{fig:descriptive} shows the histogram of the number of tweets and retweets per user for each dataset. 

The main difference between the archive accounts and timeline accounts is that the former are suspended while the latter are not. This leads to differences in characteristics of the accounts, which we will discuss in the next section. 

The datasets are made available for reproducibility\footnote{https://github.com/tugrulz/RetweetBots}.


\subsubsection{Control Groups}

\begin{figure}
    \centering
    \includegraphics[width = 0.8\columnwidth]{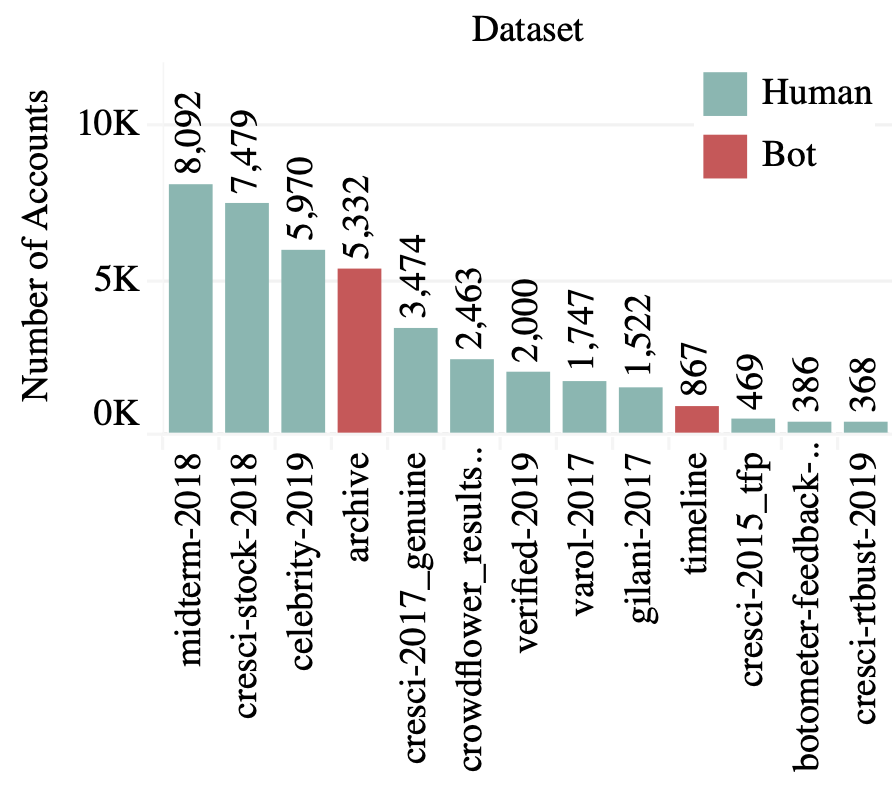}
    \caption{Dataset statistics. Each bar represents one dataset that we used in this work. The \emph{bot} bars are datasets that we built for this paper based on work by Golbeck (\citeyear{golbeck2019benford}), and the \emph{human} bars are genuine accounts from previous studies.}
    \label{fig:datasets}
\end{figure}

Understanding the differences between retweet bots and human accounts requires a control group of human (genuine) users. For this, we rely on multiple datasets of accounts that have been annotated as human-controlled in previous studies~\cite{cresci2019cashtag,yang2020scalable,mazza2019rtbust,gilani2017bots,varol2017online,cresci2017paradigm,cresci2017social,fameforsale}\footnote{https://botometer.osome.iu.edu/bot-repository/datasets.html}. We collected the timeline (most recent 3,200 tweets) of 27,622 users in 2021. The datasets' statistics are summarized in \Figref{fig:datasets}.

In some cases, our labels, which are reliable because their activity was directly purchased, do not agree with labels in other datasets. In one such dataset of human-controlled accounts~\cite{cresci2019cashtag}, which was labeled via classification, 664 and 37 bots from the timeline and archive datasets respectively are labeled as bots. However, this dataset also labels 133 bots from the timeline dataset and 1 from the archive dataset as humans. We considered these users as bots and excluded them from the humans dataset. We do not use any datasets of other types of bot accounts in the control groups because no prior dataset provided a differentiation between retweet bots and non-retweet bots. 

\subsection{Limitations}
While this dataset contains labels that are more reliable than seen in prior work, we still cannot overcome all of the limitations of data collected from the Internet. First, all data in this work came from particular black market websites. As such, we can only learn about this specific set of bots, possibly controlled by a few central points or controlled in the same manner. Other markets, or even other users on these particular markets, may have different strategies for their retweet bots which could yield different results. This is the first work to analyze retweet bots collected in this manner, so no comparisons to prior work can be made beyond those in the previous section to datasets collected using different methodologies. Still, these comparisons help us understand how different data collected in different situations can lead to different results as we will show in answering RQ4. 

Second, this dataset is still limited by the Twitter API and which data are available on the Internet Archive. These datasets do complement each other in that they contain different snapshots of Twitter, but they are still just small portions of the broader dataset of tweets. The Twitter API allows us to capture all of the \emph{not removed} content of \emph{non-suspended or private} accounts, leaving a sizable blind spot in terms of accounts that have been suspended by Twitter. This limits us in analyzing retweet bots that have purged their tweets. However, as \Figref{fig:descriptive} and \Figref{fig:activity_timeline_accounts} suggest, the majority of users have more than 1,000 retweets and were active when the retweet bots were the most active, indicating that this limitation does not harm our analysis. The API limitation of the statuses/user\_timeline endpoints also prevented us from collecting a specific user's tweets beyond the last 3,200 which affected 263 users. This is a minor issue since we were still able to collect all tweets from the majority of users and a large amount (3,200) from the remaining 263 users. The Internet Archive dataset only contains 1\% of tweets. This limits our analysis while answering RQ3 where 100\% of tweets is necessary to learn the volume of daily activity of users (A), the percentage of retweets (B), the temporal analysis (C, D), and the diversity analysis (F). 
\section{RQ1: Nature of Retweet Bots}

We first focus on meta-data analysis of the dates in which the accounts were created. We then focus on the content of the accounts and their tweets to find indicators that an account has been compromised. We find that the accounts in the archive dataset were illicitly created for this purpose, but that accounts in the timeline dataset were more likely compromised normal accounts, hence they are not suspended like the accounts in the archive dataset.

\subsection{Evidence of Mass Creation}

Many studies analyzing social media manipulation focus on accounts created around the same date and new accounts as signs of deliberate manipulation. This is based on the assumption that accounts that are created at the same time are likely to be controlled by one entity, and as such are fake (vs. either genuine or compromised). 

\begin{figure}[tb]
    \centering
    \includegraphics[width = 0.8\columnwidth]{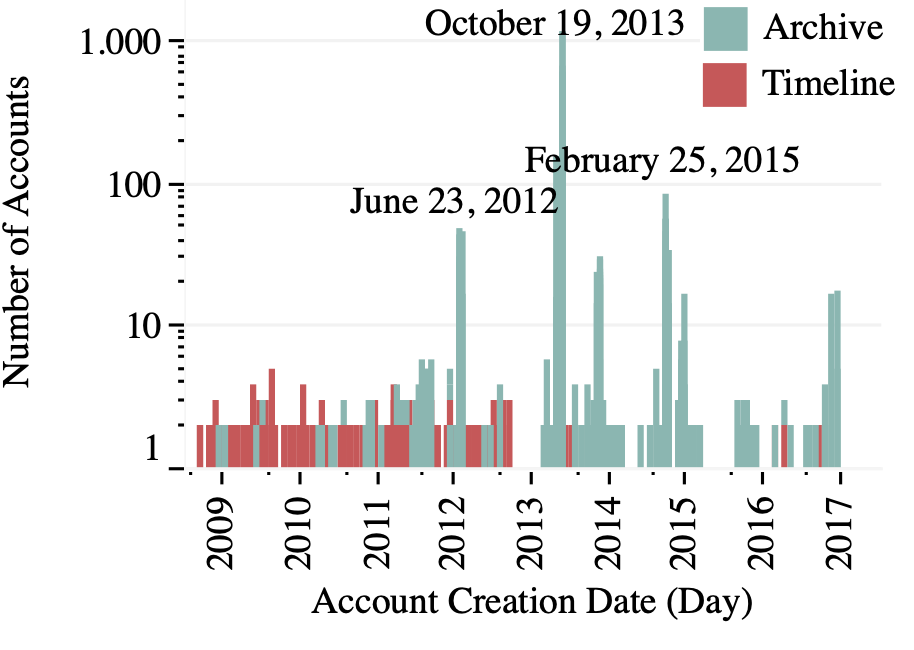}
    \caption{Creation dates of accounts in the timeline and archive datasets. Most of the accounts in the archive dataset but not the timeline dataset were created in bulk.}
    \label{fig:creation_dates}
\end{figure}

We followed this assumption and computed the number of accounts created per day. A histogram is shown in  \Figref{fig:creation_dates} to illustrate this. \emph{We found several periods in which the accounts in the archive dataset, but not in the timeline dataset, were bulk created.} The most significant period was between the 18\textsuperscript{th} and 21\textsuperscript{st} of October 2013 in which 3,750 accounts (70.3\% of accounts in this dataset) were created. 

\subsection{Evidence of Mass Compromisation}


The second potential origin of the accounts in this dataset is that they began as normal accounts and then were later compromised and used as retweet bots. While the accounts in the archive dataset were created at once, \emph{the accounts in the timeline dataset appear to be compromised.}

We find evidence of this by analyzing the \emph{content} of the tweets. That is, we compared the tweets in the timeline dataset authored during the suspected time of compromise (March-August 2017, see \Figref{fig:activity_timeline_volume}) to those authored before this period. 
There were 4,708 tweets from 184 users during the retweeting period, while there were 31,847 tweets by 322 accounts before March 2017. 

\begin{figure}[t]
    \centering
    \includegraphics[width = 0.9\columnwidth]{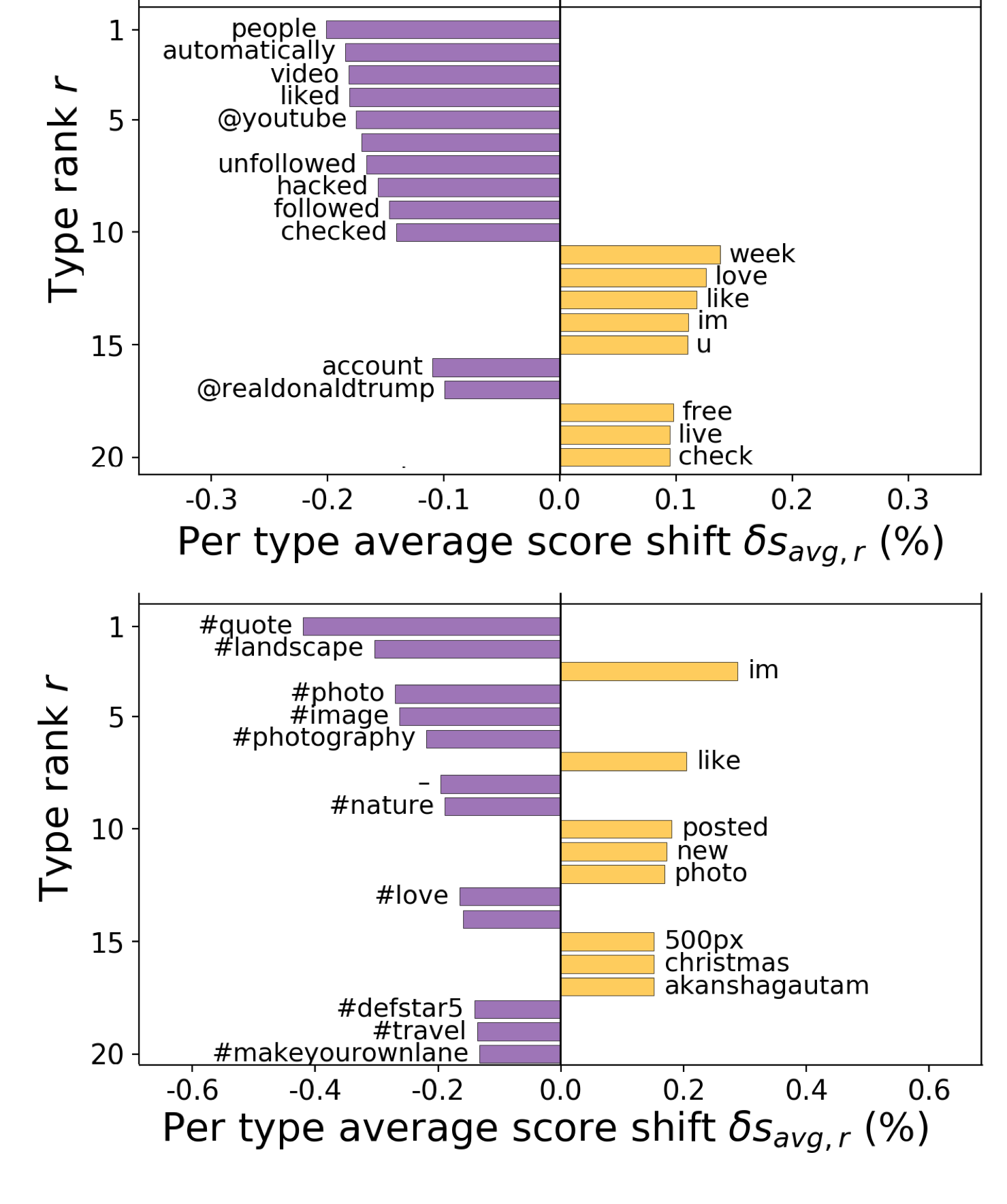}
    \caption{The wordshift graphs of accounts in the timeline dataset (upper) and archive dataset (lower). The words on the \emph{left} represent the tweets during the period of retweet activity and those on the \emph{right} represent tweets from before.} 
    \label{fig:wordshifts}
\end{figure}

To highlight the differences between the tweets in these two different periods, we use wordshift graphs~\cite{gallagher2021generalized}. Wordshift graphs compare two corpora and rank the words by their contributions to the differences in these two corpora. The contribution is computed using Shannon entropy. We randomly sample 10 tweets per account in order to give equal weights to all accounts and create two corpora, one for before the period of retweets and one during. \Figref{fig:wordshifts} shows the top words for each. We manually inspected the tweets containing these words. 

We find the most evidence from the words that were prevalent during the retweet period, but not before, in the timeline dataset (the top left of \Figref{fig:wordshifts}). Similar to prior work~\cite{zangerle2014sorry}, we found that some users were posting tweets that directly stated that their accounts had been ``hacked'' and are now recovered. We see this in the prevalence of the words \textit{hacked} and \textit{account}, as users recover their accounts over this period. The substring \textit{hack} was present in 134 tweets, and we manually verified that 42 of them (by 33 users) stated that the user's Twitter account was hacked. Some users complained about the retweets from their account, e.g., \textit{My account got \#hacked and I've tried to dlt all the retweets, but it still says I have over 2,000 tweets. How do I get rid of them? \#help.} Some users also announced that they were leaving their account because it was compromised and urged their friends to follow their new account. One user even changed their name to \textit{hacked} and changed their description to their new handle.  

Alongside \textit{hacked} and \textit{account}, we also find \textit{people}, \textit{automatically}, \textit{followed}, and \textit{unfollowed}, all words posted by a common spam app that reports how many new accounts followed/unfollowed the account owner, i.e. \textit{x people followed/unfollowed me//automatically checked by\ldots} The words \textit{video}, \textit{liked}, and \textit{@youtube} come from a Twitter app that posts the users' activity on Youtube to their Twitter feed. It is not clear if the users actually signed up for this service or if it is due to the account being compromised. The popular words \textit{love}, \textit{like}, \textit{im}, etc., distinguish the two periods. This is likely because the tweets authored by the users during the period of retweets are mainly automated messages from a script so they do not contain such otherwise popular verbs. 


We do the same analysis on the suspended users in the archive dataset. The retweet period for these accounts was longer and the edges less defined. We analyze the period in which at least half of these accounts were actively retweeting: April 2015 to October 2017. There were 10,196 tweets by 484 users before the period and 7,054 tweets by 138 users within the retweet period. Inspecting the wordshift graph, we find no patterns as we did in the timeline dataset. There are no tweets by the spam app, only two users posted Youtube activity, and we did not identify a single user complaining about getting hacked. We did find other patterns: 17 users shared quotes from famous people, and the hashtagged words (e.g. \emph{\#landscape}) were used to share a blogpost containing photos accompanied by a quote (these appear to be part of a promotion). Cresci et al. (\citeyear{cresci2017paradigm}) report a similar behavior, finding that novel social bots share quotes to appear genuine. Note that these hashtags are the most popular hashtags on Instagram\footnote{ https://influencermarketinghub.com/most-popular-instagram-hashtags} which may be the reason why these accounts target them as genuine accounts often use them as well. In conclusion, we believe that those users are fake rather than being compromised and their non-retweets are also used for promotions.

\subsection{Are any of the accounts genuine?}

Prior work~\cite{collusiveretweeters} has found that there are genuine users who sign up to blackmarket schemes to retweet others in exchange for retweets of their own content. These accounts become part of an illegitimate scheme called collusive retweeting. \emph{We find no evidence of this happening in the accounts analyzed for this paper.} Inspecting the \textit{favorite counts} and \textit{retweets counts} of the accounts' tweets, we see that most accounts do not get attention from other accounts, as would be the case if such a scheme were present in these datasets. Most of the accounts in both the timeline dataset (58\%) and the archive dataset (98.2\%) received no retweets. Even fewer accumulated between 1-10 retweets: 25.8\% in the timeline dataset and 0.5\% in the archive dataset. Finally, just 24 accounts in the timeline dataset and 36 in the archive dataset received at least 100 retweets in total. We see this mirrored for favorites counts. It is possible that the users who participated in collusive retweeting removed their retweets before data collection, therefore hiding this activity.

\section{RQ2: Lifetime of a Retweet Bot}
\label{sec:rq2}

Now that we have an understanding of where these accounts originated, we try to understand their lifespans. We investigate the time and the volume of the activity of the accounts. \emph{We observe that the accounts in our dataset were overactive in a short time period and were otherwise idle, even though they were not suspended.}

\begin{figure}[tb]
    \centering
    \includegraphics[width = 0.8\columnwidth]{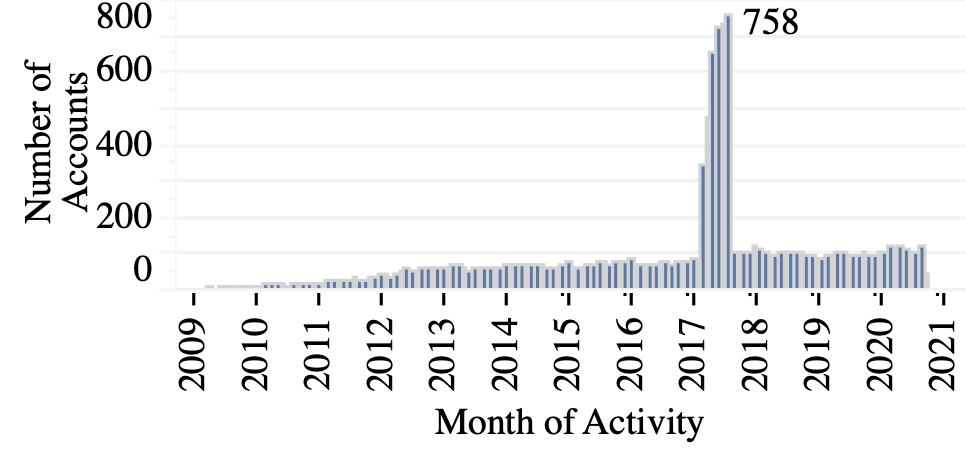}
    \caption{The number of accounts active in the timeline dataset per month. Most of the accounts were active between March 2017 and August 2017.}
    \label{fig:activity_timeline_accounts}
\end{figure}

\begin{figure}[tb]
    \centering
    \includegraphics[width = 0.8\columnwidth]{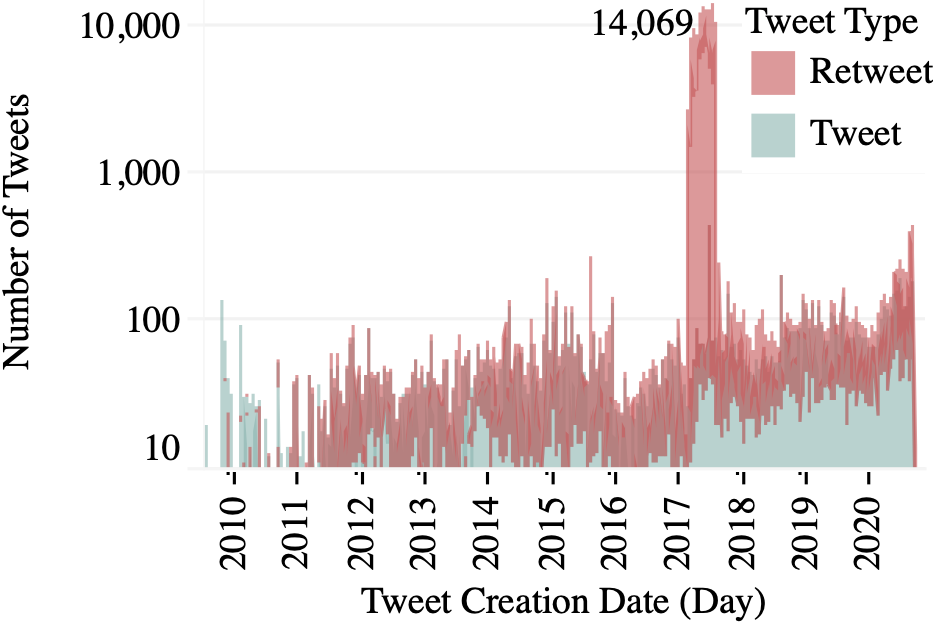}
    \caption{Number of tweets and retweets per day by the accounts in the timeline dataset. The accounts retweeted aggressively between March 2017 and August 2017, despite the low number of original tweets.}
    \label{fig:activity_timeline_volume}
\end{figure}

We found that 816 (94\%) of the accounts in the timeline dataset were active between March-August 2017, peaking in August with 758 accounts as seen in ~\Figref{fig:activity_timeline_accounts}. Note that we were not able to collect the activity of 263 accounts before March 2017 due to API limitations. The number of active accounts gradually falls to 379 on August 25th and then suddenly to 17 on August 26th. It then never exceeds 40. We make the same observation for the volume of retweets. As seen in \Figref{fig:activity_timeline_volume}, most of the retweets were posted between March 2017 and August 2017, peaking on July 30, 2017, with 14,069 retweets. The plausible explanation for this is that some malicious actor took possession of these accounts in this period and used them aggressively to retweet others. Twitter then may have imposed a ``softban'' on these accounts without actually suspending them, e.g. by asking for phone verification.

\begin{figure}[tb]
    \centering
    \includegraphics[width=0.8\columnwidth]{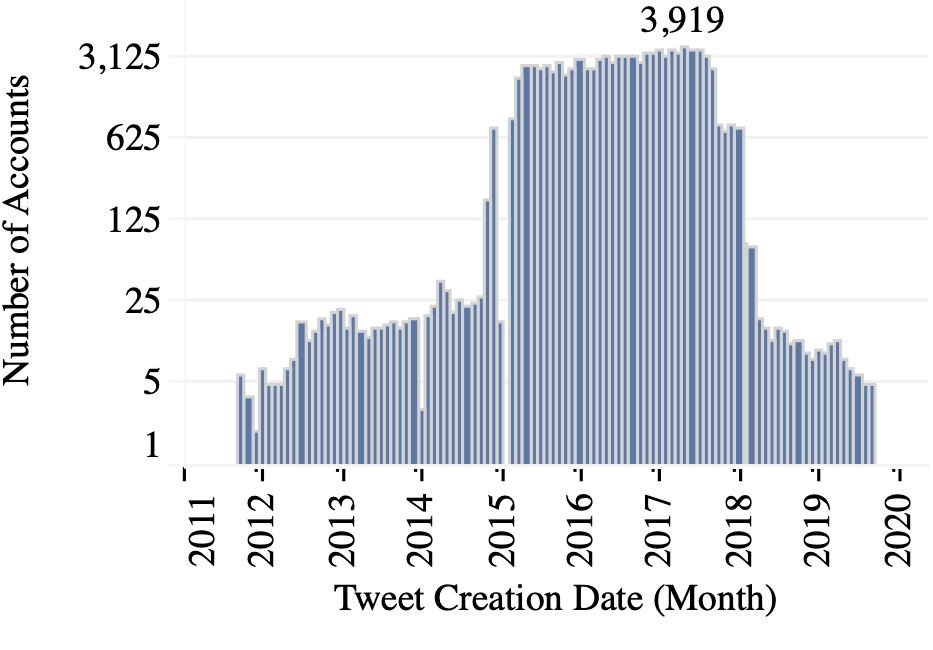}
    \caption{Number of accounts active per month. Most accounts were active between April 2015 and September 2017.}
    \label{fig:activity_archive_users}
\end{figure}

We made a similar observation with the suspended accounts from the archive dataset. As seen in \Figref{fig:activity_archive_users} the number of active accounts peak in Summer 2017, drops abruptly from 2,544 in September 2017 to 842 in October 2017 and then later to 81 in February 2018. However, the accounts were active for a longer time period than those in the timeline dataset, i.e., there were already 2,170 active accounts in April 2015. Unlike the timeline dataset accounts, these were suspended by Twitter.  What is unclear is whether they went inactive due to the suspension or whether they became inactive and were later suspended.

These results suggest that the retweets we analyzed stay active for 6 months to 2.5 years. Although this may not generalize to all retweets bots, it gives an indication of how long a retweet bot may be active.

\section{RQ3: Retweeters vs. Humans}
We next analyze how retweet bots differ from humans in terms of the time and volume of their activities. Table \ref{tab:rq3summary} summarizes the results. Specifically, we focus on the following patterns: A) volume of activity, B) percentage of retweets, C) time between consecutive retweets of the same user, D) time between consecutive retweets of the same post, E) time between the retweeted post and the retweets by the retweeter accounts F) diversity of retweeted users. The percentage of retweets and time between consecutive retweets of the same post clearly demonstrate the retweet bot activity while the other activities have patterns that challenge some of the fundamental assumptions in the bot literature. Each pattern is described by a different measurement. We compute the mean of that measure for each group (humans and bots) and report the difference in terms of that measure. We use Welch's t-test to test the statistical significance of the difference. Welch's t-test is best suited for this task as it does not assume equal variance between groups. Additionally, for patterns A and E, we compute additional measures. Two of the measures report the percentage of accounts with a binary property within each group. We report the difference in percentages and apply a chi-squared test to test the statistical significance.

\newcommand{\colwid}{90pt}

\begin{table}[ht]
\centering
\small
\begin{tabular}{p{2pt}p{\colwid}p{22pt}p{22pt}p{22pt}}
\S & Measurement & Human & Bot & Diff \\ \hline
A) & Mean status count & 6.1k & 24.5k & 18.4* \\
A) & \% of accounts with \textgreater 50 tweets/day & 42\% & 16\% & 26\%** \\
A) & Mean of max \# daily tweets & 60.6 & 40.2 & 20.4* \\
A) & Mean \# daily tweets & 9.7 & 12.2 & 2.5* \\
B) & \% of RTs & 34\% & 91\% & 57\%* \\
C) & Mean of per-user median time between RTs & 253 & 69 & 184* \\
D) & Mean of per-post mean time between RTs& 1,469 & 3.4 & 1,465.5* \\
E) & \% of RTs within 1 min.& 4\% & 1.25\% & 2.75\%** \\
E) & Mean retweet delay & 313 & 246 & 67* \\
F) & Mean Diversity & 0.50 & 0.52 & 0.2* \\ \hline\hline
\multicolumn{5}{l}{\small *p \textless 0.0001, Welch's t-test, **p \textless 0.0001, chi-squared test}   
\end{tabular}
\caption{Summary of the quantitative differences between retweeter bots and humans. All were statistically significant.}
\label{tab:rq3summary}
\end{table}

\subsection{A) Volume of Activity}

\begin{figure}[bt]
    \centering
    \includegraphics[width = \columnwidth]{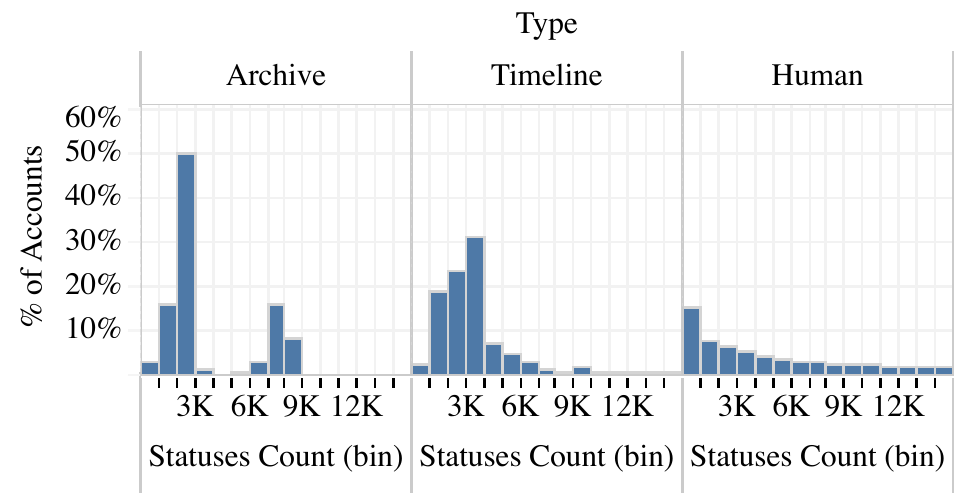}
    \caption{Histogram of status counts per day per account. While the humans’ status counts distribution follows a power law, the bots' are unimodal/bimodal, concentrated between 1,000 and 3,000, with the accounts in the archive dataset having another focal point between 7,000 and 9,000.}
    \label{fig:statuses_count}
\end{figure}

We begin by analyzing the overall activity of retweet bots and human accounts. \Figref{fig:statuses_count} shows the average daily statuses count (the total number of tweets and retweets) for each account in each dataset. Humans were more active overall, and their statuses count follows a power law. Meanwhile, the statuses counts of the bots in the timeline dataset are unimodal, and those in the archive dataset are bimodal. The latter is likely due to the presence of multiple vendors. The difference between the mean statuses count for bots (6,153) and humans (24,560) is 18,407.

We also find that the human accounts in our dataset were more likely to be overactive than bots when considering daily activity. The average number of daily tweets for bots and humans is similar, 12.2 and 9.7 respectively, however, bots tended to retweet more consistently, as we see in the variance of the number of daily tweets, 21 and 370 respectively. We see this inconsistency as well when we look at the maximum of statuses posted in a single day, i.e., the number of statuses each account posted on their most active day. Human-controlled accounts posted an average of 60.2 statuses on their most active day. This figure is 40.2 for bots.

\begin{figure}[tb]
    \centering
    \includegraphics[width = \columnwidth]{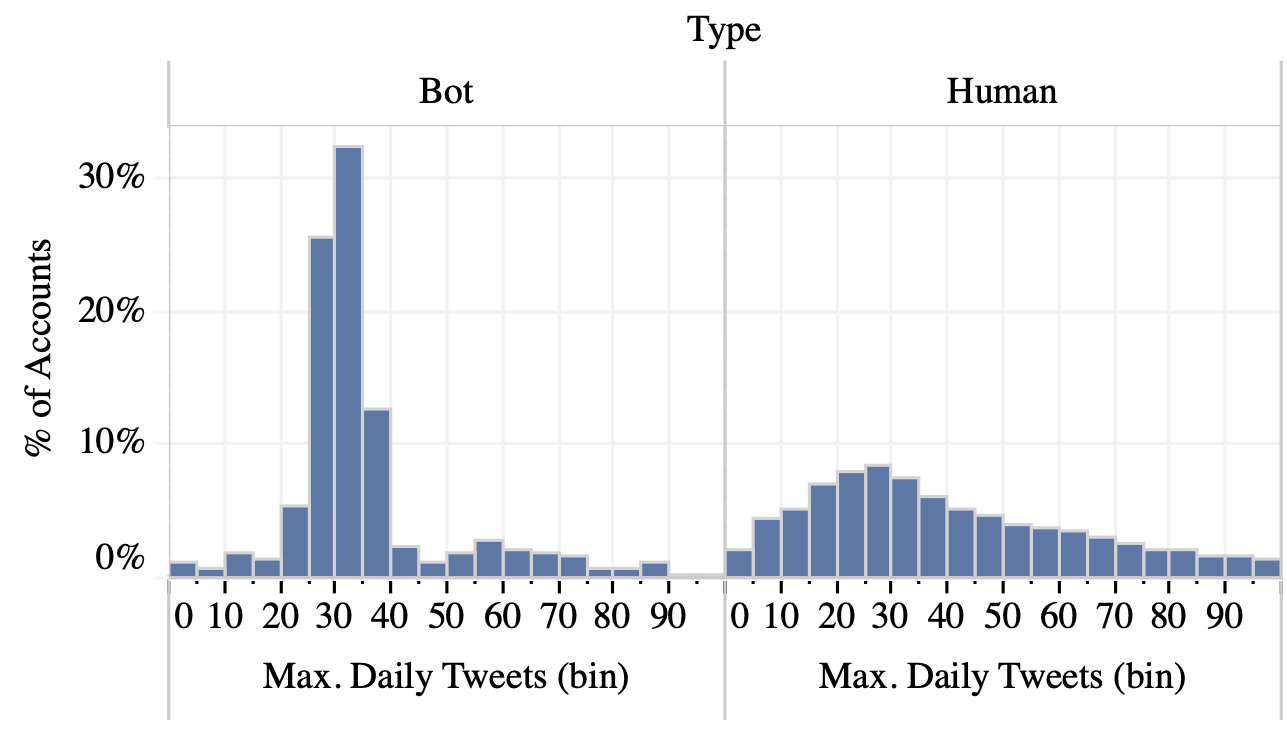}
    \caption{Maximum number of daily tweets per account. This is concentrated between 25-40 for accounts in the timeline dataset. Meanwhile, humans are more likely to be overactive and reach more than 50 tweets per day.}
    \label{fig:retweeter_overactivity}
\end{figure}

As seen in ~\Figref{fig:retweeter_overactivity}, 70\% of retweet bots had a maximum between 25-40, which is either due to a threshold set by the vendor or is constrained by the number of tasks the account vendor receives every day. Of the human accounts, 42\% tweeted at least 50 times in a single day (a threshold used by ~\cite{howard} to indicate an overactive bot), while only 16\% of retweet bots were this active. 

\subsection{B) Percentage of Retweets}

\begin{figure}[tb]
    \centering
    \includegraphics[width = 0.8\columnwidth]{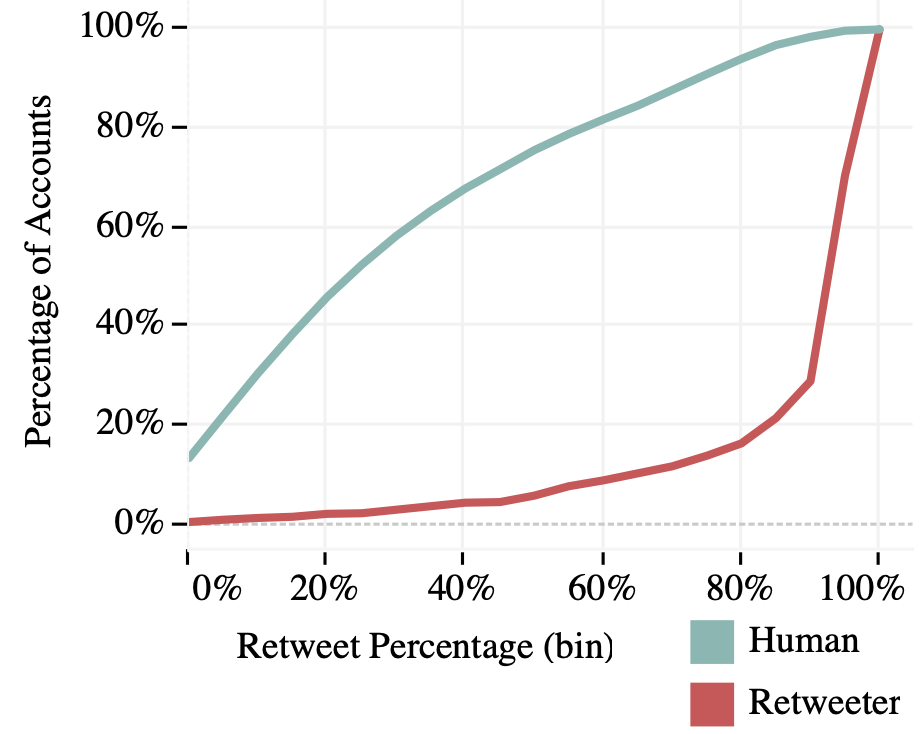}
    \caption{Cumulative frequency distribution of bots (in the timeline dataset) and human accounts according to their percentage of retweets. On average, retweeter accounts have a higher percentage of retweets.}
    \label{fig:retweeter_percentage_cdf}
\end{figure}

One major difference we expect to see with respect to these bots in particular is their retweet activity. We compute the \textit{retweet percentage} for each account, i.e., the share of retweets in all statuses the account posted. The overall average retweet percentage for bots is 91.3\% and for humans is 34.5\%.  \Figref{fig:retweeter_percentage_cdf} demonstrates how stark this difference is: 86\% of retweet bots vs 9\% of humans have a retweet percentage of at least 80\% and 95\% of retweet bots vs 28\% of humans have a retweet percentage of at least 50\%. 

Upon further inspection of the human accounts with a high retweet percentage, we found that a majority come from just two datasets: 43\% of the human accounts with a retweet ratio greater than 50\% come from cresci-stock-2018 and 22\% from midterm-2018. Additionally, cresci-rtbust-2019 contains a high percentage of such accounts. The accounts in these datasets were collected using topics that are vulnerable to manipulation (i.e. elections and financial campaigns). Midterm-2018 and cresci-rtbust-2019 were annotated by hand and cresci-stock-2018 by classification.

\subsection{C) Time Between Consecutive Retweets of the User}

\begin{figure}[tb]
    \centering
    \includegraphics[width = \columnwidth]{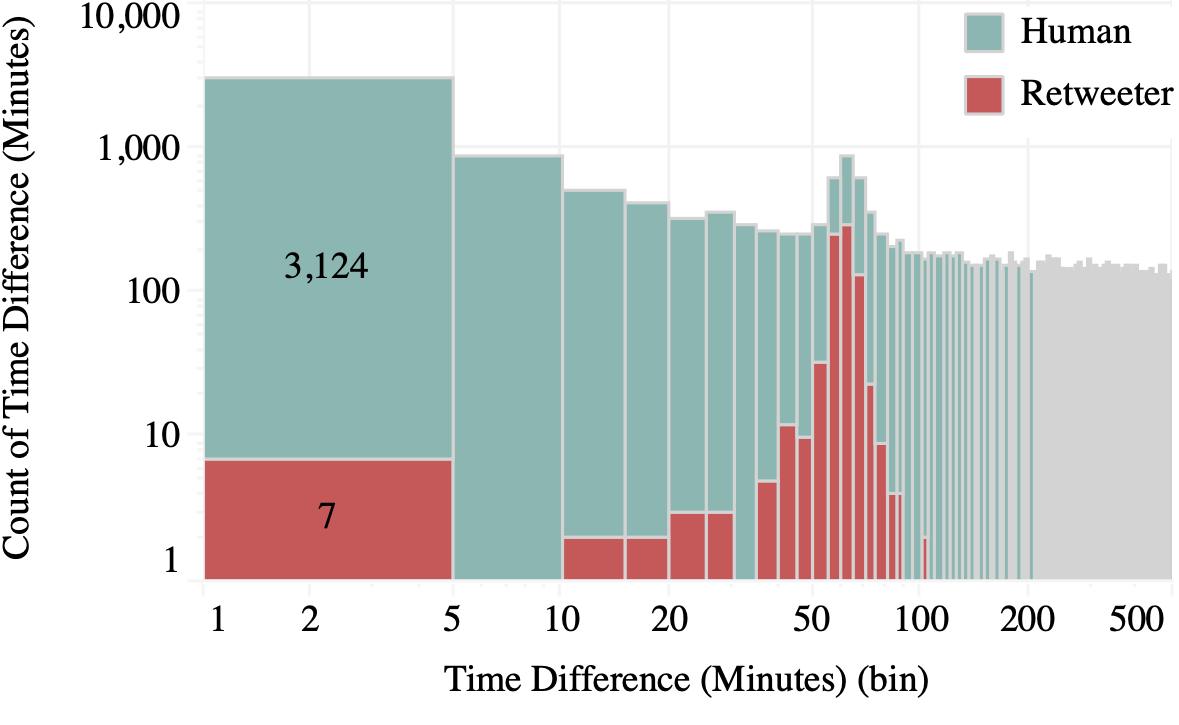}
    \caption{Median time difference between consecutive retweets per user. Retweet bots are more likely to stay idle between retweets.}
    \label{fig:consecutive_retweets_per_same_user}
\end{figure}

We further find that bots are not overactive in shorter periods, as it may be expected that they retweet in bulk and then stay idle. \Figref{fig:consecutive_retweets_per_same_user} shows the median time between consecutive retweets of each account. Bots waited roughly 60 minutes between consecutive retweets while humans tended to retweet more rapidly. Note that this is regardless of their retweet count; the bots with a very high retweet count also have the same statistic. The vendor controlling the accounts may be setting sleep times or may be limited by when the requests are made. The average of the median time differences between consecutive retweets is 69 minutes for bots and 253 minutes for humans.

\subsection{D) Time Between Consecutive Retweets per Post}

\begin{figure}[tb]
    \centering
    \includegraphics[width = 0.8\columnwidth]{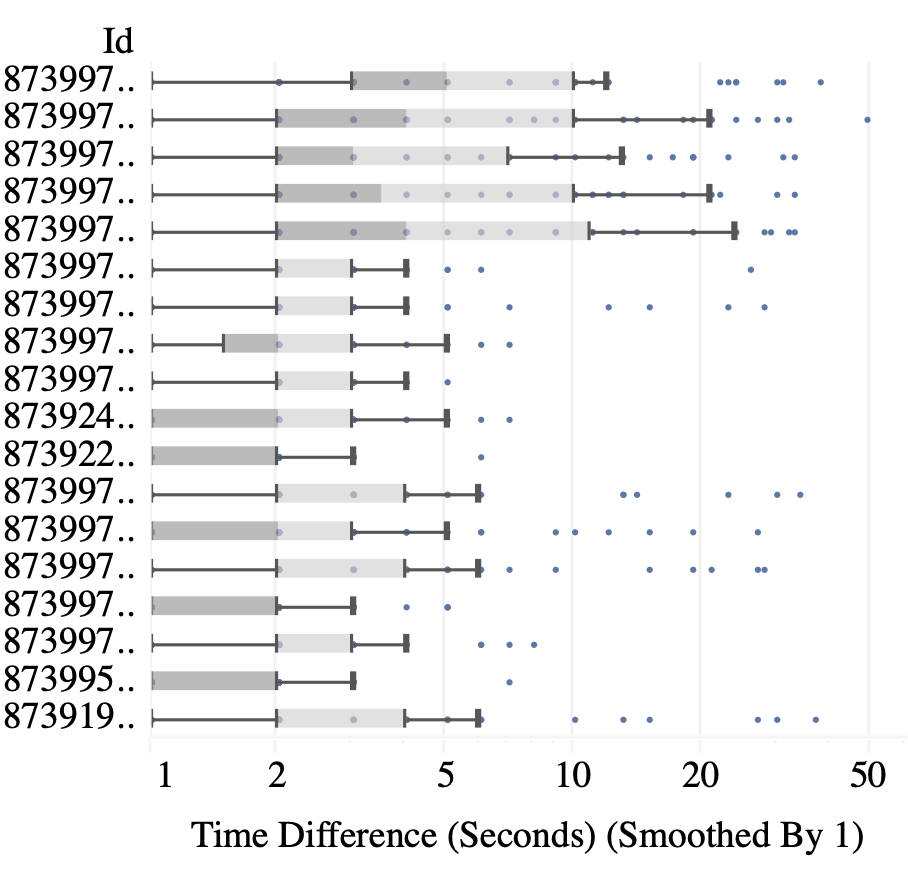}
    \caption{Time difference between consecutive retweets per post. The time differences do not exceed 50 seconds. The time difference is smoothed by 1.}
    \label{fig:consecutive_retweets_per_post_bots}
\end{figure}

Since it is the job of a retweet bot to promote posts, one might expect that while they tend to have a long delay between their own retweets, they retweet a new tweet quickly after it is posted. \Figref{fig:consecutive_retweets_per_post_bots} shows the time between each retweet of tweets that were paid to be promoted by the bots. This time difference is roughly within five seconds, and no outlier exceeds 50 seconds. 

\begin{figure}[tb]
    \centering
    \includegraphics[width = 0.9\columnwidth]{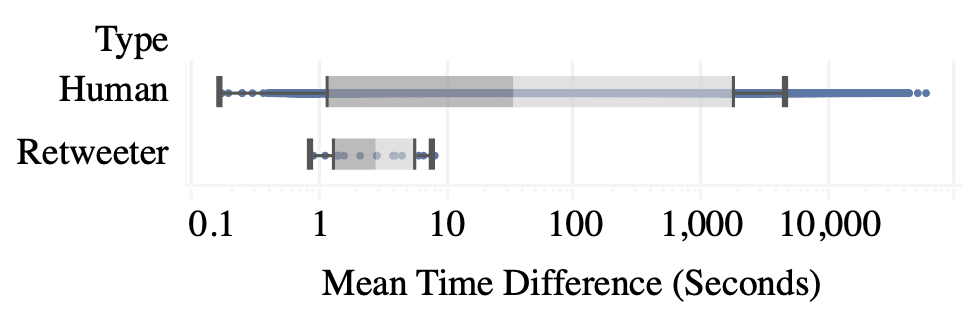}
    \caption{The mean time difference between consecutive retweets per post. It is lower for posts promoted by retweets when compared to humans.}
    \label{fig:mean_time_diff}
\end{figure}

To compare this finding with organically retweeted posts, we collected all 17,456 tweets that were retweeted by human users in our control group and had retweet counts similar to those of the bot-promoted posts (between 40 and 70). We compute the mean time between each retweet. As \Figref{fig:mean_time_diff} shows, these figures are low for both groups, however, humans have more outliers due to late retweets, so the interval of the mean time between retweets is higher. The average time between retweets is 3.4 seconds for posts promoted only by bots and 1,469 seconds for posts by humans. Retweeters retweet at roughly the same time, while humans tend to retweet with delays. 

\subsection{E) Retweet Delay}

\begin{figure}[t]
    \centering
    \includegraphics[width = \columnwidth]{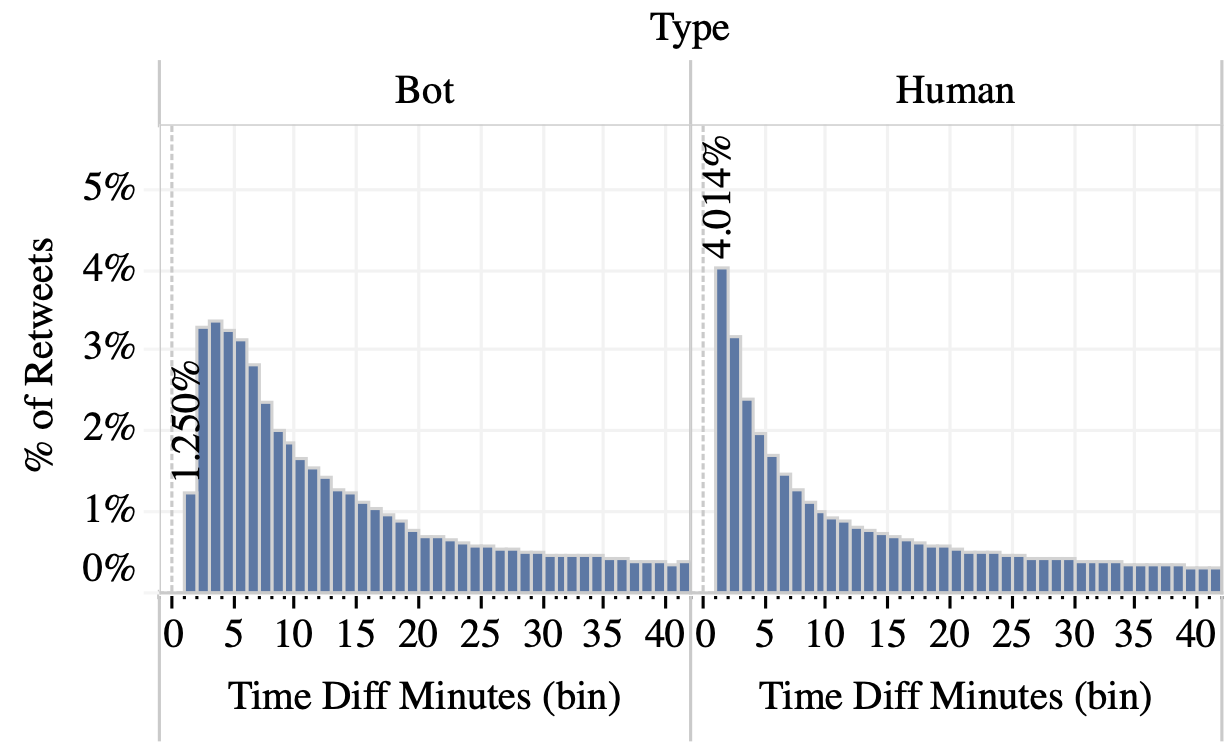}
    \caption{Time difference between the retweeted posts and retweets. Bots are more likely to have a small delay than humans who react more quickly.}
    \label{fig:time_between_original_tweet_and_retweet}
\end{figure}

Humans are able to retweet a post as quickly as they see it (e.g., after receiving a notification), but this may not be the case with retweet bots whose services are sold on the blackmarket. \Figref{fig:time_between_original_tweet_and_retweet} shows the retweet delay, i.e., the time between the tweeting being posted and the time of the retweet, for retweet by humans and bots. Humans were much more likely to retweet a tweet within the minute it was posted, while retweet bots were more likely to wait.  The accounts most likely to retweet within a minute were celebrity accounts (5.4\% of their retweets are within a minute) and verified accounts (6.4\%). These only account for 2.3\% of accounts in the archive dataset and 1\% of accounts in the timeline dataset. Overall, 4\% of humans' retweets and 1.25\% of bots' retweets were posted within a minute. 

\begin{figure}[tb]
    \centering
    \includegraphics[width = \columnwidth]{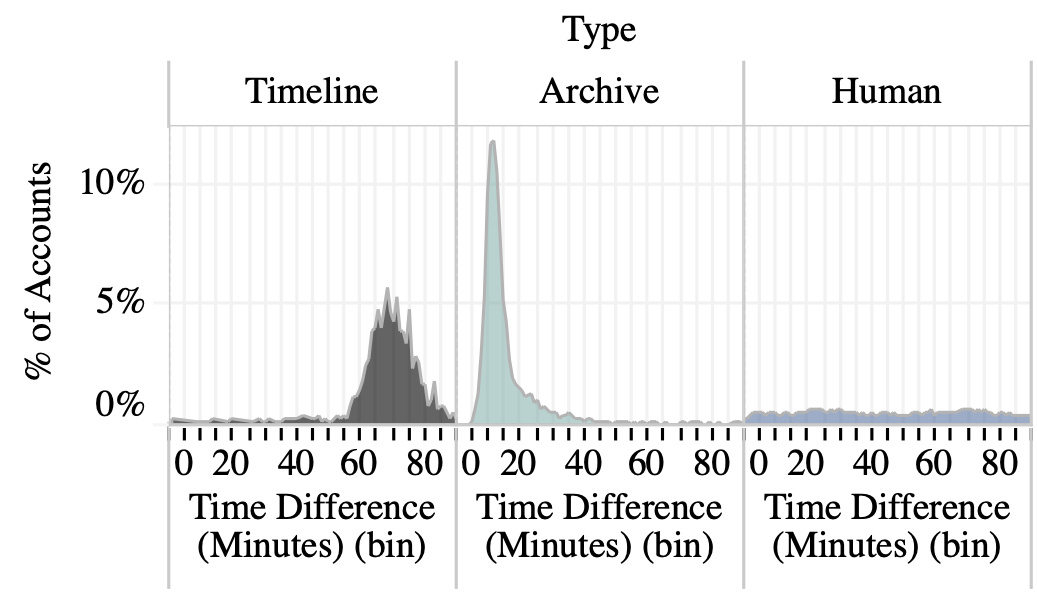}
    \caption{Median time difference between the original post and the retweet per user. It is concentrated between 8-18 minutes for accounts in the archive dataset and 60-80 minutes for accounts in the timeline dataset.}
    \label{fig:time_between_original_tweet_and_retweet_median}
\end{figure}

While the median time difference between the original post and the retweet per account is more or less uniformly distributed (but gradually declining) for humans, it is concentrated between 8-18 minutes for suspended accounts in the archive dataset and 60-80 minutes for accounts in the timeline dataset as seen in \Figref{fig:time_between_original_tweet_and_retweet_median}. Overall, bots retweet faster, with an average retweet delay of 246 minutes vs 313 minutes for humans.

These results may suggest that while humans who are active on Twitter can retweet a tweet within a minute, there is always a delay for vendor-purchased retweet bots. This is either because the vendor places a deliberate delay or because it takes at least a minute for the customer and/or the vendor to send a retweet command to retweet bots. The fact that the median time difference of retweets exceeds 8 minutes suggests that some customers do retrospective commands, i.e., they first wait for their posts to get genuine attention and only purchase retweets if they fail to.

\subsection{F) Diversity in Retweeted Users}

\begin{figure}[t]
    \centering
    \includegraphics[width=\columnwidth]{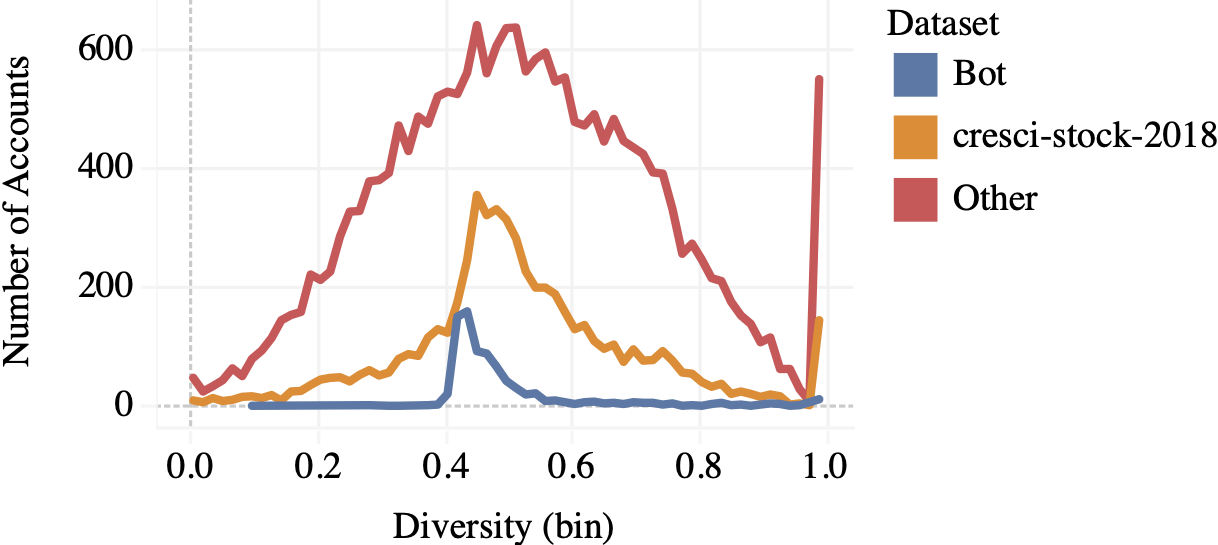}
    \caption{Number of accounts by the diversity of retweeted accounts; computed by dividing the number of unique users retweeted by the number of retweets. It follows a normal distribution for human accounts but is concentrated at 0.4-0.5 for accounts in the timeline and cresci-stock-2018 datasets.}
    \label{fig:diversity}
\end{figure}

While humans follow a set of people and, thus, are likely to retweet the same set of users, we expect retweeters to retweet from a diverse set of whichever accounts recently paid them. We compute diversity by diving the number of unique users an account retweeted by the number of retweets, shown in \Figref{fig:diversity}. We observe a normal distribution for all human datasets except for the cresci-stock-2018, which claims to be made up of human accounts with many retweets and has a distribution similar to the retweet bots (we have excluded the incorrectly included bots). 

\section{RQ4: Differences to Prior Studies}

Many bot detection methods utilize machine learning classification based on labeled datasets of bots. Often these labels come from human annotators or rely on other signals such as accounts that were banned by Twitter. Using machine learning classification for such a task has several drawbacks, but one, in particular, is that they rely on features found in training data that are inherently unstable. In this paper, we utilize datasets collected without the use of machine learning and instead rely on observing directly purchased behaviors and lateral propagation. In this section, we explore some deviations from the narratives and assumptions found in prior work that impact these features and hint at the unreliability of such classification methods.  
In some cases, these comparisons are difficult to make because many studies neglect to perform any analysis of the features used and instead rely on the classifier to determine the feature importance. That is, as long as the classifier performs well, the direction in which the feature was predictive (or even if it was predictive) is not considered. 

\subsubsection{Delayed Activity}
One common feature found in prior bot detection studies~\cite{fameforsale,varol2017online,yang2013empirical,yang2020scalable,beskow2018bot} is account age. The assumption is often that new accounts are more suspicious than older, more established ones. However, in both datasets, we found that the high activity period of the accounts occurred years after they were created. Most of the accounts in the archive dataset were created in 2013 and became active in April 2015. Meanwhile, accounts in the timeline dataset were mostly created between 2009-2013 but only became active in March 2017. Given the length of these delays, it is likely that the account owners are strategically not using freshly made accounts for malicious activity to avoid suspicion and thus detection by Twitter. This finding does not imply that creation date is not a signal for fake account detection, only that there are accounts that do not have this signal.

\subsubsection{Volume of Activity}

Some studies use the volume of activity as an indicator of fake accounts. Howard and Kollayi (\citeyear{howard}), classified overactive accounts (at least 50 tweets/day) as bots. Similarly, Dutta et al.  (\citeyear{dutta2020hawkeseye}) annotated accounts as retweeters if they retweeted many tweets in a short time. In answering RQ2, we extensively analyzed the volume of activity of the accounts in both datasets. We found that the maximum daily activity for retweeters was between 25-40, much lower than the threshold of 50, and even lower than humans. Our analysis corroborates prior work~\cite{gallwitz} claiming this assumption is faulty. 

\subsubsection{Retweet Delay} 

In answering RQ3, we learned that the bots in our dataset were on average slower or as slow as humans in retweeting a new post. Prior work~\cite{mazza2019rtbust}) introduced normal and suspicious retweeting patterns. They define a normal pattern as one with a delay based on the assumption that the fact that people see their timeline in reverse chronological order introduces delays in retweets. That is, a normal distribution centered at 100 minutes. They define a suspicious pattern in which users retweet in a matter of seconds. Our analysis contradicts this analysis, suggesting that humans are (even more) likely to retweet in seconds and their retweeting activities do not have long delays. 

\subsubsection{Diversity}
Some studies ~\cite{cresci2017social,mazza2019rtbust} find that the retweet bots they analyze, even if they take precautions to appear genuine, exclusively retweet from one or a small subset of accounts. In answering RQ3, we found that, on the contrary, they maintain a diverse set of accounts to retweet from.

\subsubsection{Friends and Followers}

\begin{figure}[bt]
    \centering
    \includegraphics[width = 0.9\columnwidth]{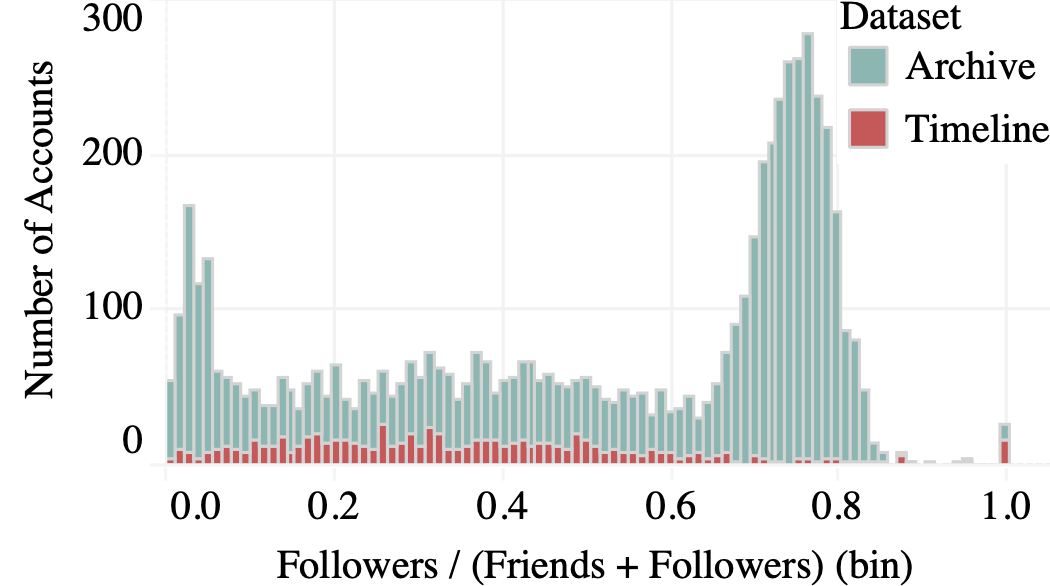}
    \caption{Follower ratios of accounts in the archive and timeline datasets. Surprisingly, accounts in the archive dataset had many more followers than friends. }
    \label{fig:friendsratio}
\end{figure}

Some prior work assumes that bots have a low follower to friend ratio because 1) they are only used to inflate follower counts of others~\cite{fameforsale}, 2) they follow many accounts in order to receive ``followbacks''~\cite{chu2010tweeting}, and/or 3) they fail to gain followers because they are illegitimate and thus have uninteresting profiles~\cite{stringhini2010detecting}. We found that 57\% (3,175) of accounts in the archive dataset and 24.4\% (212) of accounts in the timeline dataset had more followers than friends. \Figref{fig:friendsratio} shows the proportion of followers to friends for both datasets. Note that one potential source of bias in this type of ratio is accounts with very few friends and followers (e.g. an account with one friend and two followers). However, 56.8\% (3,031) of accounts have more than 100 followers. Meanwhile, we could not find a single instance of a retweet bot following another bot.

\subsubsection{Temporality} As we found while answering RQ2, the bots used for analysis in this paper are not bots at every moment of their existence. Compromised accounts began as normal accounts, became bots, and then either became normal accounts again or became inactive. Fake accounts began as inactive accounts, became bots, and then resumed their inactivity. Most bot detection works do not consider this temporality and try to detect bot activity based on the entire timeline of an account's life. For example, we found that retweet bots which were previously classified as social bots by prior work~\cite{cresci2019cashtag}, were in fact a mixture of human and inactive accounts. The accounts were compromised for only a short period, then they became inactive, possibly due to a soft ban by Twitter. It is not surprising that these accounts were classified as bots, they have an unnatural spike in the number of retweets just after compromisation, nor is it surprising that a bot detection system trained on these accounts would classify inactive accounts as bots, which was the case with Botometer~\cite{gallwitz,elmas2019lateral}. The issue comes from the underlying assumption that a bot account, by its nature, is always automated, which neglects the fact that they can be compromised and act as a bot for only a particular period of time. 
\section{Implications for Bot Detection}

\subsubsection{Feature Engineering and Overfitting}

Bot detection methods must consider the inherent assumptions made in feature engineering. The primary culprit here is that what is intuitive is not always true. For example, it may be intuitive that bot accounts follow more accounts than follow them back, but we found no evidence of this in our analysis, so a classifier trained on data biased to this assumption may not identify the bots studied in this work. 

Even in the instances when feature engineering is more quantitative than qualitative, the data itself may not be representative of bots more generally so any conclusions drawn from it about which features are important for bots at large may be inaccurate. Of course, this is always a problem for machine learning classification: when a new class or variant of an existing class is not considered in training, the classifier cannot be expected to classify it correctly. Hence, when the results of a bot classifier are used to label a dataset for analysis, these results must be considered as reflecting \emph{the types of bots that were used for training the classifier} and not bots more generally.  

To help remedy this, studies could consider focusing on bots according to their functions instead of using broad umbrella concepts like ``social bot'' which may bias the labeling and classification. In this way, we can be more accurate and precise when talking about bots and their behaviors. This can also result in more reliable account labels, as discussing bots in terms of their functions can allow for the direct purchasing of bots' activity instead of relying on humans. 

\subsubsection{Problem Setup}
Many studies focus on quantitative methods to detect and/or analyze the prevalence of bots while few focus on explaining what the detected accounts truly are. No study investigates whether the detected bots actually emulate humans. This has direct implications: the retweet bots appeared to be emulating humans because they were humans. They had full and genuine profiles because they belonged to humans once. Bot detection studies could address this using qualitative methods to determine what the accounts were truly created for in order to explain their behavior.

Similarly, we saw when studying temporality that we must consider \emph{when} an account is a bot and not only \emph{that} it is. This goes before and beyond feature engineering and into the actual problem setup and preprocessing of the training data. Studies should consider an account not as a static entity, but as something that evolves and changes throughout its life. Studies proposing new bot datasets should specify what period of activity the annotators inspected during annotation. For the data used in this paper, it is not advantageous to label retweeters in the timeline dataset as ``bots'' outside of the period of March-August 2017, before they were compromised or after they left the botnet. We already observed the implication of neglecting the temporality, as consistently classifying inactive accounts as bots is a reproducible error in Botometer~\cite{gallwitz}. 

\subsubsection{Beyond Classification}

Our analysis shows that there is a sharp distinction between retweet bots and humans with respect to retweet percentages, so a simple rule-based classifier based on retweet percentage would already suffice to classify retweet bots \emph{like the ones from this study}. However, if applied to live Twitter accounts, it would likely result in the misclassification of thousands of real users who use Twitter to amplify their favorite users, a problem already pointed out by Rauchfleisch and Kaiser (\citeyear{rauchfleisch2020false}). This is because while we know that the retweeters analyzed in this study are bots based on the fact that their activity was directly purchased, we do not know which overactive retweeter accounts are solely controlled by humans. A next step to remedy this would be to find ways to collect a dataset that represents such human-controlled accounts. 

The style of data collection used in this study is often  difficult to obtain, requires institutional backing to purchase such accounts, and results in smaller datasets than classification. In cases where we use human-labeled data or classifier outputs, either by necessity or convenience, we must consider the assumptions that went into building the datasets and the classifier and both understand and acknowledge the implications of these assumptions on our results. 

\subsubsection{Data Considerations}

Collecting data for bot studies in a way that is ethical, inexpensive, well-labeled, and large-scale is challenging. There is often a trade-off between the certainty of labels, ease of collection (and therefore scale of the dataset), and an ethical data collection. The dataset used in this study sacrifices size and collection ease in favor of label certainty. In doing so, we found that there were some assumptions and findings from prior work that did not hold in our dataset. Future research should consider these trade-offs carefully in light of such findings.

\section{Ethics}

Collecting high-quality data to study bots is challenging, but this work demonstrates the necessity of correctly labeling accounts as bots or humans.  As such, we must be able to label some bots based on a third channel of information (e.g. purchasing). This introduces some ethical issues that do not arise when hand labeling or using classification, as in most cases this involves participation in the black market. 

To weigh the ethics of using data collected in a way that could be unethical, we refer to recent work by Ienca and Vayena (\citeyear{Ienca2021ethical}) on responsibly using hacked data for research. While our work is not a straightforward application of their arguments, it is nonetheless a useful tool for understanding the ethics of using third-party datasets that may have been collected either unethically or in breach of the Terms of Service of the platform. This work argues that one must weigh the public value, optimization of resources, uniqueness, and cross-domain consistency of the leaked data (in our case the archive dataset) against consent issues, possible secondary harms, breach of privacy, and lower quality data of an otherwise non-leaked dataset (a hand-labeled dataset collected using only the Twitter API). In this paper, we opted to not actively participate in the black market, but instead use data that others had collected from blackmarket sources. In this way, we mitigate the harms of further participation in the market. This also mitigates another ethical dilemma: hand-labeling is by its nature error-prone and, thus, inevitably leads to human accounts being labeled as bots, as we found in answering \emph{RQ4}. This can cause harms to these users who are treated as bot accounts. Finally, these data are also very unique and without them, we are unable to claim a reliable ground truth.

In terms of data collection, throughout this work, we did not collect any data from Twitter except by use of the Twitter API, and with this data, we respect the rules and regulations written in the Twitter Terms of Service. We do utilize a dataset that was released by The Internet Archive, which may have been collected in breach of the Twitter terms of service because it contains deleted and suspended content. Referring again to the work of Ienca and Vayena, we weigh the utility and uniqueness of using deleted content collected by a third party, some of which is content that Twitter removed due to inauthentic bot activity and some of which was deleted by the bots to hide their activity~\cite{elmas2020power}, against the privacy of users who have deleted their content since the data was collected by The Internet Archive. Indeed, bot research has an immense value for the public good as bots spread fake news and promote harmful narratives. 

\section{Conclusion}

While we are beginning to understand how spam bots, fake follower accounts, and astroturfing bots operate, retweet bots have been largely unexplored. In this work, we presented a dataset of retweet bots that were directly purchased from vendors on the black market and analyzed them for the purpose of learning where they originate, how long they are active, and how they differ from human accounts. This unique dataset gives us new insights into the world of bots. In studying the behavior and lifespan of retweet bots, we also found several inconsistencies between our results and those obtained using bots studied in prior works. These results challenge some of the basic understandings about bot behavior and operation and highlight the need for studies that use reliable datasets and make decisions about bots based on them.

\fontsize{9.0pt}{10.0pt} \selectfont
\bibliography{bib}

\begin{thebibliography}{39}
\providecommand{\natexlab}[1]{#1}
\providecommand{\url}[1]{\texttt{#1}}
\providecommand{\urlprefix}{URL }
\expandafter\ifx\csname urlstyle\endcsname\relax
  \providecommand{\doi}[1]{doi:\discretionary{}{}{}#1}\else
  \providecommand{\doi}{doi:\discretionary{}{}{}\begingroup
  \urlstyle{rm}\Url}\fi

\bibitem[{Archive(2020)}]{team2020archive}
Archive. 2020.
\newblock The twitter stream grab. Accessed on 2020-12-01.

\bibitem[{Beskow and Carley(2018)}]{beskow2018bot}
Beskow, D.~M.; and Carley, K.~M. 2018.
\newblock Bot-hunter: a tiered approach to detecting \& characterizing
  automated activity on twitter.
\newblock In \emph{Conference paper. SBP-BRiMS: International Conference on
  Social Computing, Behavioral-Cultural Modeling and Prediction and Behavior
  Representation in Modeling and Simulation}.

\bibitem[{Boshmaf et~al.(2011)Boshmaf, Muslukhov, Beznosov, and
  Ripeanu}]{boshmaf2011socialbot}
Boshmaf, Y.; Muslukhov, I.; Beznosov, K.; and Ripeanu, M. 2011.
\newblock The socialbot network: when bots socialize for fame and money.
\newblock In \emph{Proceedings of the 27th annual computer security
  applications conference}.

\bibitem[{Chavoshi, Hamooni, and Mueen(2016)}]{debot}
Chavoshi, N.; Hamooni, H.; and Mueen, A. 2016.
\newblock DeBot: Twitter Bot Detection via Warped Correlation.
\newblock In \emph{ICDM}.

\bibitem[{Chu et~al.(2010)Chu, Gianvecchio, Wang, and
  Jajodia}]{chu2010tweeting}
Chu, Z.; Gianvecchio, S.; Wang, H.; and Jajodia, S. 2010.
\newblock Who is tweeting on Twitter: human, bot, or cyborg?
\newblock In \emph{Proceedings of the 26th annual computer security
  applications conference}.

\bibitem[{Chu, Widjaja, and Wang(2012)}]{spamcampaign}
Chu, Z.; Widjaja, I.; and Wang, H. 2012.
\newblock Detecting social spam campaigns on twitter.
\newblock In \emph{International Conference on Applied Cryptography and Network
  Security}.

\bibitem[{Cresci et~al.(2015)Cresci, Di~Pietro, Petrocchi, Spognardi, and
  Tesconi}]{fameforsale}
Cresci, S.; Di~Pietro, R.; Petrocchi, M.; Spognardi, A.; and Tesconi, M. 2015.
\newblock Fame for sale: Efficient detection of fake Twitter followers.
\newblock \emph{Decision Support Systems} .

\bibitem[{Cresci et~al.(2017{\natexlab{a}})Cresci, Di~Pietro, Petrocchi,
  Spognardi, and Tesconi}]{cresci2017paradigm}
Cresci, S.; Di~Pietro, R.; Petrocchi, M.; Spognardi, A.; and Tesconi, M.
  2017{\natexlab{a}}.
\newblock The paradigm-shift of social spambots: Evidence, theories, and tools
  for the arms race.
\newblock In \emph{Proceedings of the 26th international conference on world
  wide web companion}.

\bibitem[{Cresci et~al.(2017{\natexlab{b}})Cresci, Di~Pietro, Petrocchi,
  Spognardi, and Tesconi}]{cresci2017social}
Cresci, S.; Di~Pietro, R.; Petrocchi, M.; Spognardi, A.; and Tesconi, M.
  2017{\natexlab{b}}.
\newblock Social fingerprinting: detection of spambot groups through
  DNA-inspired behavioral modeling.
\newblock \emph{IEEE Transactions on Dependable and Secure Computing} .

\bibitem[{Cresci et~al.(2019{\natexlab{a}})Cresci, Lillo, Regoli, Tardelli, and
  Tesconi}]{cresci2019cashtag}
Cresci, S.; Lillo, F.; Regoli, D.; Tardelli, S.; and Tesconi, M.
  2019{\natexlab{a}}.
\newblock Cashtag piggybacking: Uncovering spam and bot activity in stock
  microblogs on Twitter.
\newblock \emph{ACM Transactions on the Web (TWEB)} .

\bibitem[{Cresci et~al.(2019{\natexlab{b}})Cresci, Petrocchi, Spognardi, and
  Tognazzi}]{cresciinferhumans}
Cresci, S.; Petrocchi, M.; Spognardi, A.; and Tognazzi, S. 2019{\natexlab{b}}.
\newblock On the capability of evolved spambots to evade detection via genetic
  engineering.
\newblock \emph{Online Social Networks and Media} .

\bibitem[{Dutta, Arora, and Chakraborty(2021)}]{dutta2021abome}
Dutta, H.~S.; Arora, U.; and Chakraborty, T. 2021.
\newblock ABOME: A Multi-platform Data Repository of Artificially Boosted
  Online Media Entities.
\newblock \emph{arXiv preprint arXiv:2103.15250} .

\bibitem[{Dutta et~al.(2018)Dutta, Chetan, Joshi, and
  Chakraborty}]{collusiveretweeters}
Dutta, H.~S.; Chetan, A.; Joshi, B.; and Chakraborty, T. 2018.
\newblock Retweet us, we will retweet you: Spotting collusive retweeters
  involved in blackmarket services.
\newblock In \emph{IEEE/ACM International Conference on Advances in Social
  Networks Analysis and Mining}.

\bibitem[{Dutta et~al.(2020)Dutta, Dutta, Adhikary, and
  Chakraborty}]{dutta2020hawkeseye}
Dutta, H.~S.; Dutta, V.~R.; Adhikary, A.; and Chakraborty, T. 2020.
\newblock HawkesEye: Detecting fake retweeters using Hawkes process and topic
  modeling.
\newblock \emph{IEEE Transactions on Information Forensics and Security} .

\bibitem[{Elmas, Overdorf, and Aberer(2021)}]{elmas2021dataset}
Elmas, T.; Overdorf, R.; and Aberer, K. 2021.
\newblock A Dataset of State-Censored Tweets.
\newblock In \emph{Proceedings of the International AAAI Conference on Web and
  Social Media}, volume~15, 1009--1015.

\bibitem[{Elmas et~al.(2020)Elmas, Overdorf, Akg{\"u}l, and
  Aberer}]{elmas2020misleading}
Elmas, T.; Overdorf, R.; Akg{\"u}l, {\"O}.~F.; and Aberer, K. 2020.
\newblock Misleading repurposing on twitter.
\newblock \emph{arXiv preprint arXiv:2010.10600} .

\bibitem[{Elmas et~al.(2019)Elmas, Overdorf, {\"O}zkalay, and
  Aberer}]{elmas2019lateral}
Elmas, T.; Overdorf, R.; {\"O}zkalay, A.~F.; and Aberer, K. 2019.
\newblock Lateral Astroturfing Attacks on Twitter Trending Topics.
\newblock \emph{arXiv preprint arXiv:1910.07783} .

\bibitem[{Elmas et~al.(2021)Elmas, Overdorf, {\"O}zkalay, and
  Aberer}]{elmas2020power}
Elmas, T.; Overdorf, R.; {\"O}zkalay, A.~F.; and Aberer, K. 2021.
\newblock Ephemeral astroturfing attacks: The case of fake twitter trends.
\newblock In \emph{2021 IEEE European Symposium on Security and Privacy
  (EuroS\&P)}, 403--422. IEEE.

\bibitem[{Ferrara et~al.(2016)Ferrara, Varol, Davis, Menczer, and
  Flammini}]{ferrara2016rise}
Ferrara, E.; Varol, O.; Davis, C.; Menczer, F.; and Flammini, A. 2016.
\newblock The rise of social bots.
\newblock \emph{Communications of the ACM} .

\bibitem[{Gallagher et~al.(2021)Gallagher, Frank, Mitchell, Schwartz, Reagan,
  Danforth, and Dodds}]{gallagher2021generalized}
Gallagher, R.~J.; Frank, M.~R.; Mitchell, L.; Schwartz, A.~J.; Reagan, A.~J.;
  Danforth, C.~M.; and Dodds, P.~S. 2021.
\newblock Generalized word shift graphs: a method for visualizing and
  explaining pairwise comparisons between texts.
\newblock \emph{EPJ Data Science} .

\bibitem[{Gallwitz and Kreil(2021)}]{gallwitz}
Gallwitz, F.; and Kreil, M. 2021.
\newblock The Rise and Fall of `Social Bot' Research.
\newblock \emph{SSRN: https://ssrn.com/abstract=3814191} .

\bibitem[{Giatsoglou et~al.(2015)Giatsoglou, Chatzakou, Shah, Faloutsos, and
  Vakali}]{giatsoglou2015retweeting}
Giatsoglou, M.; Chatzakou, D.; Shah, N.; Faloutsos, C.; and Vakali, A. 2015.
\newblock Retweeting activity on twitter: Signs of deception.
\newblock In \emph{Pacific-Asia Conference on Knowledge Discovery and Data
  Mining}. Springer.

\bibitem[{Gilani et~al.(2017)Gilani, Farahbakhsh, Tyson, Wang, and
  Crowcroft}]{gilani2017bots}
Gilani, Z.; Farahbakhsh, R.; Tyson, G.; Wang, L.; and Crowcroft, J. 2017.
\newblock Of bots and humans (on twitter).
\newblock In \emph{Proceedings of the 2017 IEEE/ACM International Conference on
  Advances in Social Networks Analysis and Mining 2017}.

\bibitem[{Golbeck(2019)}]{golbeck2019benford}
Golbeck, J. 2019.
\newblock Benford’s Law can detect malicious social bots.
\newblock \emph{First Monday} .

\bibitem[{Gupta, Kumaraguru, and Chakraborty(2019)}]{gupta2019malreg}
Gupta, S.; Kumaraguru, P.; and Chakraborty, T. 2019.
\newblock Malreg: Detecting and analyzing malicious retweeter groups.
\newblock In \emph{Proceedings of the ACM India Joint International Conference
  on Data Science and Management of Data}.

\bibitem[{Herzallah, Faris, and Adwan(2018)}]{herzallah2018feature}
Herzallah, W.; Faris, H.; and Adwan, O. 2018.
\newblock Feature engineering for detecting spammers on Twitter: Modelling and
  analysis.
\newblock \emph{Journal of Information Science} .

\bibitem[{Howard and Kollanyi(2016)}]{howard}
Howard, P.~N.; and Kollanyi, B. 2016.
\newblock Bots,\# strongerin, and\# brexit: Computational propaganda during the
  uk-eu referendum.
\newblock \emph{Available at SSRN 2798311} .

\bibitem[{Ienca and Vayena(2021)}]{Ienca2021ethical}
Ienca, M.; and Vayena, E. 2021.
\newblock Ethical requirements for responsible research with hacked data.
\newblock \emph{Nature Machine Intelligence} 3(9): 744--748.

\bibitem[{Liu, Hooi, and Faloutsos(2017)}]{liu2017holoscope}
Liu, S.; Hooi, B.; and Faloutsos, C. 2017.
\newblock Holoscope: Topology-and-spike aware fraud detection.
\newblock In \emph{Proceedings of the 2017 ACM on Conference on Information and
  Knowledge Management}.

\bibitem[{Mazza et~al.(2019)Mazza, Cresci, Avvenuti, Quattrociocchi, and
  Tesconi}]{mazza2019rtbust}
Mazza, M.; Cresci, S.; Avvenuti, M.; Quattrociocchi, W.; and Tesconi, M. 2019.
\newblock Rtbust: Exploiting temporal patterns for botnet detection on twitter.
\newblock In \emph{Proceedings of the 10th ACM Conference on Web Science}.

\bibitem[{Rauchfleisch and Kaiser(2020)}]{rauchfleisch2020false}
Rauchfleisch, A.; and Kaiser, J. 2020.
\newblock The False positive problem of automatic bot detection in social
  science research.
\newblock \emph{Berkman Klein Center Research Publication} .

\bibitem[{Stringhini, Kruegel, and Vigna(2010)}]{stringhini2010detecting}
Stringhini, G.; Kruegel, C.; and Vigna, G. 2010.
\newblock Detecting spammers on social networks.
\newblock In \emph{Proceedings of the 26th annual computer security
  applications conference}.

\bibitem[{Tekumalla, Asl, and Banda(2020)}]{tekumalla2020mining}
Tekumalla, R.; Asl, J.~R.; and Banda, J.~M. 2020.
\newblock Mining Archive. org’s twitter stream grab for pharmacovigilance
  research gold.
\newblock In \emph{Proceedings of the International AAAI Conference on Web and
  Social Media}.

\bibitem[{Vargas, Emami, and Traynor(2020)}]{vargas2020detection}
Vargas, L.; Emami, P.; and Traynor, P. 2020.
\newblock On the detection of disinformation campaign activity with network
  analysis.
\newblock In \emph{Proceedings of the 2020 ACM SIGSAC Conference on Cloud
  Computing Security Workshop}.

\bibitem[{Varol et~al.(2017)Varol, Ferrara, Davis, Menczer, and
  Flammini}]{varol2017online}
Varol, O.; Ferrara, E.; Davis, C.; Menczer, F.; and Flammini, A. 2017.
\newblock Online human-bot interactions: Detection, estimation, and
  characterization.
\newblock In \emph{Proceedings of the international AAAI conference on web and
  social media}.

\bibitem[{Yang, Harkreader, and Gu(2013)}]{yang2013empirical}
Yang, C.; Harkreader, R.; and Gu, G. 2013.
\newblock Empirical evaluation and new design for fighting evolving twitter
  spammers.
\newblock \emph{IEEE Transactions on Information Forensics and Security} .

\bibitem[{Yang et~al.(2020)Yang, Varol, Hui, and Menczer}]{yang2020scalable}
Yang, K.-C.; Varol, O.; Hui, P.-M.; and Menczer, F. 2020.
\newblock Scalable and generalizable social bot detection through data
  selection.
\newblock In \emph{Proceedings of the AAAI Conference on Artificial
  Intelligence}.

\bibitem[{Yardi et~al.(2010)Yardi, Romero, Schoenebeck
  et~al.}]{yardi2010detecting}
Yardi, S.; Romero, D.; Schoenebeck, G.; et~al. 2010.
\newblock Detecting spam in a twitter network.
\newblock \emph{First Monday} .

\bibitem[{Zangerle and Specht(2014)}]{zangerle2014sorry}
Zangerle, E.; and Specht, G. 2014.
\newblock " Sorry, I was hacked" a classification of compromised twitter
  accounts.
\newblock In \emph{Proceedings of the 29th annual acm symposium on applied
  computing}.

\end{thebibliography}

\end{document}